\documentclass[5p,preprint,number,sort&compress,a4paper]{elsarticle}
\usepackage{amsmath}
\usepackage{amsfonts}
\usepackage{amssymb}
\usepackage{accents}
\usepackage{mathrsfs}
\usepackage[normalem]{ulem}
\usepackage[T1]{fontenc}
\usepackage[utf8]{inputenc}
\usepackage[colorlinks=true]{hyperref}

\newcommand{\dd}{\mathrm{d}}
\newcommand{\lc}[1]{\accentset{\circ}{#1}}
\newcommand{\torvec}{\mathscr{V}}
\newcommand{\toraxi}{\mathscr{A}}
\DeclareMathOperator{\curl}{curl}

\journal{Physics Letters B}

\begin{document}

\begin{frontmatter}
\title{Gravitational wave birefringence in spatially curved teleparallel cosmology}

\author[1]{Manuel Hohmann\corref{cor1}}
\ead{manuel.hohmann@ut.ee}

\author[2]{Christian Pfeifer}
\ead{christian.pfeifer@zarm.uni-bremen.de}

\cortext[cor1]{Corresponding author}
\address[1]{Laboratory of Theoretical Physics, Institute of Physics, University of Tartu, W. Ostwaldi 1, 50411 Tartu, Estonia}
\address[2]{Center Of Applied Space Technology And Microgravity - ZARM, University of Bremen, Am Fallturm 2, 28359 Bremen, Germany}

\begin{abstract}
We study tensor perturbations around the most general cosmologically symmetric backgrounds in a class of teleparallel gravity theories known as New General Relativity. These theories comprise a one-parameter class, which is fully consistent with observations at the post-Newtonian level, and which contains the teleparallel equivalent of General Relativity as a special case. We find that for a particular class of cosmological background geometries these theories exhibit gravitational wave birefringence and dispersion, i.e., the dispersion relation for gravitational waves depends on their polarization and wave number. The strength of this effect is directly related to the spatial curvature of the cosmological background and the parameter describing the deviation of the theory from General Relativity. We discuss the possibility of observing this effect in gravitational wave experiments.
\end{abstract}

\begin{keyword}
gravitational waves; teleparallel gravity; tensor perturbations; birefringence
\end{keyword}
\end{frontmatter}


\section{Introduction}\label{sec:intro}
Numerous observations in cosmology, as well as its tension with quantum theory, hint towards the necessity to extend general relativity (GR) by more general means than adding a cosmological constant $\Lambda$ and cold dark matter (CDM), which are the ingredients of the $\Lambda$CDM model. A large number of such extended theories of gravity has been studied under theoretical and observational as well as classical and quantum aspects~\cite{CANTATA:2021ktz,Addazi:2021xuf}. With the advent of gravitational wave (GW) observations~\cite{LIGOScientific:2016aoc}, the properties of GW have become an important observational discriminator to test the viability of gravity theories. Most prominently, the observation of the nearly simultaneous arrival of GW and an electromagnetic (EM) counterpart from a binary neutron star merger~\cite{LIGOScientific:2017vwq,LIGOScientific:2017zic} has set narrow bounds on the propagation speed of GW compared to that of light, which has ruled out several classes of gravity theories~\cite{Ezquiaga:2017ekz}. Even without the existence of an EM counterpart, a frequency-dependent propagation speed of GW can leave an imprint on the observed GW waveforms, and can thus be constrained by the more abundant observations of binary black hole mergers~\cite{Baker:2022rhh}.

A different potential discriminator is GW birefringence, i.e., a difference in the speed of propagation between different propagation eigenstates~\cite{Ezquiaga:2020dao}. The nature and cause of GW birefringence can be different, depending on the gravity theory under consideration. The most well-known case is the existence of massive and massless scalar GW modes in addition to the massless tensor modes in theories with non-minimally coupled, massive scalar fields~\cite{Liang:2017ahj,Hou:2017bqj}. A less prominent possibility is birefringence between the two different helicity eigenstates of a tensorial GW: this effect occurs typically due to the presence of parity-violating interactions in the action, possibly coupled to a scalar field, that can for example be caused by non-commutative fields or Lorentz violating operators, whose non-vanishing background value leads to a parity-violating term in the perturbed gravitational field equations which govern the GW propagation~\cite{Lue:1998mq,Cai:2007xr,Kostelecky:2016kfm,Qiao:2019wsh,Zhao:2019xmm,Li:2020xjt}.

In this letter we study another possibility to obtain birefringence between the two helicities of tensor GW, which appears even without introducing a parity-violating term into the action of the theory. The class of gravity theories we study for this purpose is known as teleparallel gravity~\cite{Aldrovandi:2013wha,Krssak:2018ywd,Bahamonde:2021gfp}. Its distinguishing feature is the presence of a flat, metric-compatible connection, whose torsion is employed as the mediator of gravity, instead of the curvature of the Levi-Civita connection as in GR. Numerous works have studied the propagation of gravitational waves in teleparallel gravity on Minkowski~\cite{GarciadeAndrade:2002mw,Bamba:2013ooa,Farrugia:2018gyz,Hohmann:2018xnb,Hohmann:2018jso,Capozziello:2020vil} and flat Friedmann-Lema\^itre-Robertson-Walker (FLRW) backgrounds~\cite{Bahamonde:2019ipm,Bahamonde:2021dqn}, also in the effective field theory approach \cite{Cai:2018rzd,Li:2018ixg}, while perturbations of spatially curved FLRW backgrounds have only recently received attention~\cite{Hohmann:2020vcv,Bahamonde:2022ohm}.

A particularly interesting teleparallel generalization of general relativity, since it has passed all viability tests so far, is a 1-parameter subclass of the 3-parameter class of theories called New General Relativity (NGR)~\cite{Hayashi:1979qx,Hayashi:1981qx}. In this letter we study its predictions on the propagation of tensor perturbations on a general homogeneous and isotropic background. It has been shown that this choice of cosmological background allows for a non-vanishing axial torsion, which can be expressed by a non-vanishing, timelike pseudo-vector \(\mathfrak{a}_{\mu}\), and is closely linked to the spatial curvature of the background~\cite{Hohmann:2020zre}. We will see that in NGR there exists a non-trivial coupling between the background axial torsion and the tensor perturbations. We will calculate their dispersion relation, which predicts GW birefringence and dispersion and give prospects on the detectability of the effect. We like to point out that in the studies of gravitational waves in $f(T)$-gravity, the mostly studied teleparallel theory of gravity in the literature, no birefringence effects have been found \cite{Cai:2018rzd,Li:2018ixg}, which is the distinguished feature we find in our analysis presented here for 1-parameter NGR (1PNGR).


We use the following conventions: the components of the Minkowski metric are $\eta_{\mu\nu} = \textrm{diag}(-1,1,1,1)$; $A,B,\ldots = 0,1,2,3$ label Lorentz indices, $\mu,\nu,\ldots = 0,1,2,3$ label spacetime indices and $a,b,\ldots = 1,2,3$ label spatial spacetime indices. We use natural units in which the speed of light is \(c = 1\).

\section{New General Relativity}
Teleparallel theories of gravity are formulated in terms of a tetrad $\theta^A{}_\mu$, which defines the spacetime metric $g_{\mu\nu}=\eta_{AB}\theta^A{}_\mu \theta^B{}_\nu$, and a flat, metric compatible connection $\omega^A{}_{B\mu}$ with torsion, which ensures local Lorentz invariance of the theory \cite{Aldrovandi:2013wha,Krssak:2018ywd,Bahamonde:2021gfp}. In the following we will work w.l.o.g., see \cite{Krssak:2015oua,Hohmann:2017duq,Blixt:2019mkt,Blixt:2022rpl,Hohmann:2021dhr}, in the Weitzenb\"ock gauge $\omega^A{}_{B\mu}=0$. In this gauge the torsion tensor, from which the gravitational field action is constructed in teleparallel gravity, is determined by the tetrad $T^\rho{}_{\mu\nu} = 2 e_A{}^\rho\partial_{[\mu}\theta^A{}_{\nu]}$. There exist three quadratic, parity even scalars
\begin{equation}\label{eq:tseven}
T_{\text{vec}} = \mathfrak{v}_{\mu}{\mathfrak{v}}^{\mu}\,, \quad
T_{\text{axi}} = \mathfrak{a}_{\mu}{\mathfrak{a}}^{\mu}\,, \quad
T_{\text{ten}} = \mathfrak{t}_{\lambda\mu\nu}{\mathfrak{t}}^{\lambda\mu\nu}\,,
\end{equation}
that can be constructed from the vector, axial, tensor decomposition of the torsion
\begin{subequations}
\begin{align}
\mathfrak{v}_{\mu} &= T^{\nu}{}_{\nu\mu}\,,\label{eq:vecT}\\
\mathfrak{a}_{\mu} &= \frac{1}{6}\epsilon_{\mu\nu\rho\sigma}T^{\nu\rho\sigma}\,,\label{eq:axT}\\
\mathfrak{t}_{\mu\nu\rho} &= T_{(\mu\nu)\rho} + \frac{1}{3}\left(T^{\sigma}{}_{\sigma(\mu}g_{\nu)\rho} - T^{\sigma}{}_{\sigma\rho}g_{\mu\nu}\right)\,.\label{eq:tenT}
\end{align}
\end{subequations}
The most general teleparallel theory of gravity, whose action can be constructed linearly from these torsion scalars has been named New General Relativity (NGR) \cite{Hayashi:1979qx,Hayashi:1981qx}, and its action reads
\begin{equation}\label{eq:gravact}
S_{\text{g}}[\theta] = \frac{1}{2\kappa^2}\int\dd^4x\,\theta(c_t T_{\text{ten}} + c_v T_{\text{vec}} + c_a T_{\text{axi}})\,,
\end{equation}
where \(c_t, c_v, c_a\) are constant parameters, and \(\theta = \det\theta^A{}_{\mu}\). This general three parameter theory possesses mathematical and physical problems such as the existence of ghost modes and constraints from the PPN analysis \cite{Hayashi:1979qx,Hayashi:1981qx}. However, for specific values of the three parameters of the theory NGR is nothing but the teleparallel equivalent of general relativity (TEGR) \cite{Maluf:2013gaa}, which is equivalent to general relativity at the level of the metric field equation and its solutions. This observation opens the question whether there exist further special values of $c_t$, $c_v$ and $c_a$ such that the aforementioned problems are avoided, but still the theory is an extension of general relativity.

The aforementioned question can be answered affirmatively. Fixing the theory parameters to their TEGR value up to a deviation $\zeta$ in $c_a$,
\begin{equation}\label{eq:onepar}
c_a = \frac{3}{2} + \zeta\,, \quad
c_v = -\frac{2}{3}\,, \quad
c_t = \frac{2}{3}\,,
\end{equation}
gives 1-parameter NGR (1PNGR) \cite{Hayashi:1979qx,Hayashi:1981qx}: a basically unconstrained teleparallel extension of general relativity whose field equations can be written as~\cite[Eq.~(5.32)]{Bahamonde:2021gfp}
\begin{equation}
\begin{aligned}\label{eq:ngrfield}
\lc{G}_{\mu\nu} + \zeta \Big(& \tfrac{1}{2} \mathfrak{a}^\sigma \mathfrak{a}_{(\sigma}g_{\mu\nu)} - \tfrac{4}{9} \epsilon_{\nu\rho\sigma\lambda} \mathfrak{a}^\rho \mathfrak{t}_{\mu}{}^{\sigma\lambda}\\
&- \tfrac{2}{9}\epsilon_{\mu\nu\rho\sigma} (\mathfrak{a}^\rho \mathfrak{v}^\sigma + \tfrac{3}{2} \lc{\nabla}^\rho \mathfrak{a}^\sigma) \Big)=\kappa^2\Theta_{\mu\nu}\,,
\end{aligned}
\end{equation}
where \(\lc{G}_{\mu\nu}\) is the Einstein tensor obtained from the metric \(g_{\mu\nu}\) and $\Theta_{\mu\nu}$ is the Hilbert energy-momentum tensor of source matter. This theory is known to propagate no other than the two tensor polarizations of gravitational waves on a flat Minkowski background, whose speed equals that of light~\cite{Hohmann:2018xnb,Hohmann:2018jso}. Further, its post-Newtonian limit agrees with that of general relativity~\cite{Ualikhanova:2019ygl}, and it has also been demonstrated that Schwarzschild geometry is a vacuum solution of the theory for arbitrary $\zeta$ \cite{Hayashi:1979qx,Hayashi:1981qx}.

Note that for \(\zeta = 0\), the action and field equations reduce to that of TEGR. In this letter we investigate the imprints of a non-vanishing value of $\zeta$ on the predictions of propagation of gravitational waves from cosmological perturbation theory.

\section{Cosmologically symmetric background}
In the following, we will assume that the teleparallel geometry, which constitutes the dynamic field of 1PNGR, is given by a small perturbation around a cosmologically symmetric, i.e., homogeneous and isotropic background geometry. The latter has been found and classified in~\cite{Hohmann:2019nat,Hohmann:2020zre} and is characterized by a FLRW metric
\begin{multline}
\bar{g}_{\mu\nu}\dd x^{\mu} \otimes \dd x^{\nu} = a(t)^2\bigg[-\dd t \otimes \dd t + \frac{\dd r \otimes \dd r}{1 - u^2r^2}\\
+ r^2(\dd\vartheta \otimes \dd\vartheta + \sin^2\vartheta\dd\varphi \otimes \dd\varphi)\bigg]
\end{multline}
using conformal time \(t\) and spherical coordinates \((r, \vartheta, \varphi)\), where the curvature parameter \(u=\sqrt{K}\) can be real or imaginary. We can split this metric in the form
\begin{equation}
\bar{g}_{\mu\nu} = -n_{\mu}n_{\nu} + h_{\mu\nu}\,,
\end{equation}
where the unit normal covector field is given by
\begin{equation}
n_{\mu}\dd x^{\mu} = -a(t)\dd t\,,
\end{equation}
using conformal time \(t\). It enables us to construct a spatial Levi-Civita tensor
\begin{equation}
\varepsilon_{\mu\nu\rho} = n^{\sigma}\bar{\epsilon}_{\sigma\mu\nu\rho}\,, \quad
\bar{\epsilon}_{\mu\nu\rho\sigma} = 4\varepsilon_{[\mu\nu\rho}n_{\sigma]}\,.
\end{equation}
from the full spacetime Levi-Civita tensor \(\bar{\epsilon}_{\mu\nu\rho\sigma}\) of \(\bar{g}_{\mu\nu}\). The cosmological teleparallel geometry is completed by the unique flat, metric-compatible connection, whose torsion tensor is given by
\begin{equation}
\bar{T}^{\mu}{}_{\nu\rho} = 2\frac{\torvec}{a}h^{\mu}_{[\nu}n_{\rho]} + 2\frac{\toraxi}{a}\varepsilon^{\mu}{}_{\nu\rho}\,,
\end{equation}
introducing the parameter functions \(\torvec = \torvec(t)\) and \(\toraxi = \toraxi(t)\). In order to ensure the flatness of the connection, they must take one of the two forms
\begin{equation}\label{eq:vectorbranch}
\torvec = \mathcal{H} \pm iu\,, \quad \toraxi = 0\,,
\end{equation}
which we will call the vector branch, or
\begin{equation}\label{eq:axialbranch}
\torvec = \mathcal{H}\,, \quad \toraxi = \pm u\,,
\end{equation}
called the axial branch. Here \(\mathcal{H} = a'/a\) is the conformal Hubble parameter, and the prime \(\prime\) denotes the derivative with respect to the conformal time \(t\). Since we assume all dynamical fields to be real, we choose \(iu \in \mathbb{R}\) for the vector branch, so that the spatial curvature parameter satisfies \(K = u^2 \leq 0\), while for the axial branch we set \(u \in \mathbb{R}\), hence \(K = u^2 \geq 0\). For \(u \to 0\), both branches assume a common limit
\begin{equation}\label{eq:flatlimit}
\torvec = \mathcal{H}\,, \quad \toraxi = 0\,.
\end{equation}
Inserting this background geometry into the gravitational field equations~\eqref{eq:ngrfield} of 1PNGR, we find
\begin{subequations}\label{eq:cosmobgfield}
\begin{align}
3\torvec(2\mathcal{H} - \torvec) + (3 + 2\zeta)\toraxi^2 &= \kappa^2a^2\bar{\rho}\,,\\
\torvec(\torvec - 2\mathcal{H}) - 2\torvec' - \left(1 + \frac{2}{3}\zeta\right)\toraxi^2 &= \kappa^2a^2\bar{p}
\end{align}
\end{subequations}
for the two branches given above, where the background energy-momentum tensor is given by the perfect fluid form
\begin{equation}
\bar{\Theta}_{\mu\nu} = \bar{\rho}n_{\mu}n_{\nu} + \bar{p}h_{\mu\nu}
\end{equation}
in terms of the density \(\bar{\rho}\) and pressure \(\bar{p}\).

\section{Tensor perturbations}\label{sec:tenpert}
In order to discuss tensor perturbations and their dynamics, we write the spatial part of the background geometry in terms of a conformally rescaled time independent metric~$\gamma_{ab}$ defined as
\begin{equation}
h_{\mu\nu}\dd x^{\mu} \otimes \dd x^{\nu} = a^2\gamma_{ab}\dd x^a \otimes \dd x^b\,.
\end{equation}
Further, we need its Levi-Civita tensor $v_{abc}$, defined by
\begin{equation}
\varepsilon_{\mu\nu\rho}\dd x^{\mu} \otimes \dd x^{\nu} \otimes \dd x^{\rho} = a^3\upsilon_{abc}\dd x^a \otimes \dd x^b \otimes \dd x^c\,,
\end{equation}
its Levi-Civita covariant derivative \(\dd_a\) and the corresponding Laplacian \(\triangle = \dd_a\dd^a\).  Latin indices denote spatial coordinates, which are raised and lowered with \(\gamma_{ab}\). Following the procedure detailed in~\cite{Hohmann:2020vcv}, the perturbations of the cosmologically symmetric background metric and teleparallel connection are governed by a linear perturbation \(\tau_{\mu\nu}\) of the tetrad which defines the background geometry. This yields the perturbed metric
\begin{equation}
g_{\mu\nu} = \bar{g}_{\mu\nu} + 2\tau_{(\mu\nu)}\,,
\end{equation}
while the perturbed flat, metric-compatible connection becomes
\begin{equation}
\Gamma^{\mu}{}_{\nu\rho} = \bar{\Gamma}^{\mu}{}_{\nu\rho} + \bar{\nabla}_{\rho}\tau^{\mu}{}_{\nu}\,.
\end{equation}
From now on we raise and lower indices with the background metric, as usual in linear perturbation theory. In the following we study the symmetric trace- and divergence-free tensorial part $q_{\mu\nu}$ of the general perturbations $\tau_{\mu\nu}$, so that the latter takes the form
\begin{equation}
\tau_{\mu\nu} = \frac{1}{2}q_{\mu\nu}\,.
\end{equation}
We rescale these perturbations conveniently,
\begin{equation}
q_{\mu\nu}\dd x^{\mu} \otimes \dd x^{\nu} = a^2\hat{q}_{ab}\dd x^a \otimes \dd x^b\,,
\end{equation}
and the rescaled quantities satisfy
\begin{equation}
\hat{q}_{[ab]} = 0\,, \quad
\hat{q}_a{}^a = 0\,, \quad
\dd^a\hat{q}_{ab} = 0\,.
\end{equation}
In addition, we consider a perturbation of the energy-momentum tensor of the analogous form
\begin{equation}
\Theta_{\mu\nu} = \bar{\Theta}_{\mu\nu} + \mathcal{T}_{\mu\nu}\,,
\end{equation}
where the relevant part of the anisotropic stress \(\mathcal{T}_{\mu\nu}\) which sources the dynamics of the perturbations \(q_{\mu\nu}\) is the symmetric trace- and divergence-free part. Defining
\begin{equation}
\mathcal{T}_{\mu\nu}\dd x^{\mu} \otimes \dd x^{\nu} = a^2\hat{\mathcal{T}}_{ab}\dd x^a \otimes \dd x^b\,,
\end{equation}
one has
\begin{equation}
\hat{\mathcal{T}}_{[ab]} = 0\,, \quad
\hat{\mathcal{T}}_a{}^a = 0\,, \quad
\dd^a\hat{\mathcal{T}}_{ab} = 0\,.
\end{equation}
Inserting this tensor perturbation of the cosmologically symmetric geometry and energy-momentum into the gravitational field equations~\eqref{eq:ngrfield}, and assuming that the background equations~\eqref{eq:cosmobgfield} are satisfied, one finds that the only non-vanishing perturbed field equation has the same tensorial character as the perturbations. For the axial branch~\eqref{eq:axialbranch}, it reads
\begin{multline}\label{eq:tenspertaxi}
\hat{q}_{ab}'' + 2\mathcal{H}\hat{q}_{ab}' - \triangle\hat{q}_{ab} + \left(2 - \frac{4}{3}\zeta\right)u^2\hat{q}_{ab} + \frac{4}{3}\zeta u\dd_c\hat{q}_{d(a}\upsilon_{b)}{}^{cd}\\
= 2\kappa^2a^2\hat{\mathcal{T}}_{ab}\,,
\end{multline}
while for the vector branch~\eqref{eq:vectorbranch}, one finds
\begin{equation}\label{eq:tenspertvec}
\hat{q}_{ab}'' + 2\mathcal{H}\hat{q}_{ab}' - \triangle\hat{q}_{ab} + 2u^2\hat{q}_{ab} = 2\kappa^2a^2\hat{\mathcal{T}}_{ab}\,.
\end{equation}
In the spatially flat limit \(u \to 0\), both reduce to
\begin{equation}\label{eq:tenspertflat}
\hat{q}_{ab}'' + 2\mathcal{H}\hat{q}_{ab}' - \triangle\hat{q}_{ab} = 2\kappa^2a^2\hat{\mathcal{T}}_{ab}\,.
\end{equation}
In order to further analyze the solutions to these equations, we assume vanishing anisotropic stress, \(\hat{\mathcal{T}}_{ab} = 0\), and expand the geometry perturbation \(\hat{q}_{ab}\) into tensor harmonics \(\hat{e}^{\pm}_{ab}(\beta)\). The latter can most easily be defined by introducing the curl
\begin{equation}
\curl\hat{q}_{ab} = \upsilon_{cd(a}\dd^c\hat{q}_{b)}{}^d\,,
\end{equation}
which turns out to satisfy
\begin{equation}
\curl\curl\hat{q}_{ab} = -\triangle\hat{q}_{ab} + 3u^2\hat{q}_{ab}\,.
\end{equation}
The tensor harmonics come in two helicities \(\hat{e}^{\pm}_{ab}(\beta)\) satisfying
\begin{equation}
\curl\hat{e}^{\pm}_{ab}(\beta) = \pm\beta\hat{e}^{\pm}_{ab}(\beta)\,,
\end{equation}
where \(\beta \in \{3, 4, 5, \ldots\}\) for \(u^2 = 1\) and \(\beta \geq 0\) otherwise~\cite{Abbott:1986ct}. It follows that they are eigenfunctions of the Laplacian with eigenvalue
\begin{equation}
k^2\hat{e}^{\pm}_{ab}(\beta) = -\triangle\hat{e}^{\pm}_{ab}(\beta) = (\beta^2 - 3u^2)\hat{e}^{\pm}_{ab}(\beta)\,.
\end{equation}
We can then expand the geometry perturbation in the form
\begin{equation}
\hat{q}_{ab} = \sum_{I,\beta,\omega}Q^I(\beta, \omega)\hat{e}^I_{ab}(\beta)e^{-i\omega t}\,,
\end{equation}
where the symbol \(\sum\) is supposed to denote a sum in case of a discrete parameter range, and an integral in the continuous case. For a single mode, the axial branch field equation~\eqref{eq:tenspertaxi} then yields the dispersion relation
\begin{equation}\label{eq:dispaxi}
\omega^2 + 2i\mathcal{H}\omega = \beta^2 - \left(1 + \frac{4}{3}\zeta\right)u^2 \pm \frac{4}{3}\zeta u\beta\,,
\end{equation}
while for the vector branch~\eqref{eq:tenspertvec} we have
\begin{equation}\label{eq:dispvec}
\omega^2 + 2i\mathcal{H}\omega = \beta^2 - u^2\,,
\end{equation}
with
\begin{equation}
\beta = \sqrt{k^2 + 3u^2}\,.
\end{equation}
Hence, we see that while the dispersion relation~\eqref{eq:dispvec} for the vector branch does neither depend on the helicity, nor on the parameter $\zeta$, it receives a contribution depending on both $\zeta$ and the helicity in the axial branch~\eqref{eq:dispaxi}. The latter therefore leads to a birefringence of gravitational waves, whose strength depends on the curvature parameter \(u\) as well as the 1PNGR parameter~\(\zeta\).

\section{Experimental bounds}
In the following, we will restrict ourselves to the axial branch~\eqref{eq:axialbranch}, which is the only branch for which gravitational wave birefringence occurs. Further, in the short wavelength case \(|\omega| \gg |\mathcal{H}|\), we can neglect the Hubble friction term. From the dispersion relation~\eqref{eq:dispaxi}, we then obtain the phase velocity 
\begin{equation}
c_p^{\pm} = \frac{\omega}{k} = \sqrt{1 + \frac{2u}{3k^2}\left[u(3 - 2\zeta) \pm 2\zeta\sqrt{k^2 + 3u^2}\right]}\,,
\end{equation}
as well as the group velocity 
\begin{equation}
c_g^{\pm} = \frac{\dd\omega}{\dd k} = \frac{1}{c_p^{\pm}}\left(1 \pm \frac{2u\zeta}{3\sqrt{k^2 + 3u^2}}\right)\,,
\end{equation}
relative to an observer who is comoving with the cosmological FLRW background. Similarly, one could derive $c_g^{\pm}$ by interpreting~\eqref{eq:dispaxi} as the definition of a Hamiltonian via $\mathfrak{H}(t,\omega, k)=0$, then $c_g^{\pm}=\partial_k \mathfrak{H}/\partial_{\omega}\mathfrak{H}$, a feature often used to derive time delays from modified dispersion relations~\cite{Pfeifer:2018pty}. Note that both velocities depend on the wave number \(k\) and the helicity, the latter being encoded in the sign \(\pm\). Thus we find birefringence as well as dispersion of the gravitational wave.

We then calculate the relative deviation
\begin{equation}
\Delta c_{p,g} = \frac{c_{p,g}^+ - c_{p,g}^-}{c_{p,g}^+ + c_{p,g}^-}\,.
\end{equation}
Performing a Taylor expansion in the curvature parameter \(u\), we find that the leading terms are given by
\begin{equation}
\Delta c_p \approx \frac{2u\zeta}{3k}\,, \quad
\Delta c_g \approx -\frac{2u^3\zeta(3 + 2\zeta)^2}{27k^3}\,.
\end{equation}
In order to compare these results to experiments and obtain bounds on the model parameters, we finally need to convert the appearing quantities into physical units. For this purpose, we introduce the curvature density parameter
\begin{equation}
\Omega = -\frac{u^2}{\mathcal{H}_0^2} = -\frac{u^2}{a_0^2H_0^2}\,,
\end{equation}
as well as the physical wave number
\begin{equation}
\tilde{k} = \frac{k}{a_0}\,,
\end{equation}
where the subscript \(0\) in both equations denotes the value at the present time, while \(H = \mathcal{H}/a\) is the cosmological Hubble parameter. In the following, we will neglect the redshift of the source, as it is not essential for a first order approximation. Replacing \(u\) and \(k\) with \(\Omega\) and \(\tilde{k}\) in the formulas above, one finds that the scale factor \(a_0\) cancels as expected and one obtains the physical values
\begin{subequations}
\begin{align}
\Delta c_p &\approx \frac{2H_0\zeta\sqrt{-\Omega}}{3\tilde{k}}\,,\\
\Delta c_g &\approx -\frac{2H_0^3\zeta(3 + 2\zeta)^2\sqrt{-\Omega^3}}{27\tilde{k}^3}\,.
\end{align}
\end{subequations}
Note that we would have obtained the same result as the leading terms in a Taylor expansion with respect to \(H_0/\tilde{k}\) from the full expression. Using the approximate values \(H_0 \approx 68\mathrm{km/s/Mpc}\) and \(\tilde{k}/(2\pi) \approx 100\mathrm{Hz}\) for a typical gravitational wave signal originating from a stellar binary black hole merger, one arrives at \(H_0/\tilde{k} \approx 3.5 \cdot 10^{-21}\), which justifies considering only these leading terms. To link these results to particular observations, we first consider the induced phase shift. For a signal emitted at comoving distance \(d\) at a constant frequency \(\omega\), the phase at the observer is
\begin{equation}
\Phi = dk^{\pm} = \frac{d\omega}{c_p^{\pm}}\,.
\end{equation}
Using physical units \(\tilde{d} = a_0d\) and \(\tilde{\omega} = \omega/a_0\), the phase shift between the two helicities to first order is given by
\begin{equation}
\Delta\Phi \approx -\tilde{d}\tilde{\omega}\frac{c_p^+ - c_p^-}{(c_p^{\pm})^2} \approx -2\tilde{d}\tilde{k}\Delta c_p \approx -\frac{4\tilde{d}H_0\zeta\sqrt{-\Omega}}{3}\,.
\end{equation}
For a source at a distance of \(\tilde{d} = 1\mathrm{Gpc}\), one has \(\tilde{d}H_0 \approx 0.23\).
Further, for a signal originating from a singular event we have the arrival time difference
\begin{equation}
\Delta t \approx -\tilde{d}\frac{c_g^+ - c_g^-}{(c_g^{\pm})^2} \approx -2\tilde{d}\Delta c_g \approx \frac{4\tilde{d}H_0^3\zeta(3 + 2\zeta)^2\sqrt{-\Omega^3}}{27\tilde{k}^3}\,.
\end{equation}
For the numerical values we assumed above and using suitable units, we find the factor \(H_0^3/\tilde{k}^3 \approx 4.3 \cdot 10^{-62} \approx 4.4 \cdot 10^{-45}\mathrm{s/Gpc}\), and so this effect is beyond the bounds of detectability, unless \(\zeta\) takes very high values. This holds also in the LISA band \(\tilde{k}/(2\pi) \approx 10^{-3}\mathrm{Hz}\), which yields \(H_0^3/\tilde{k}^3 \approx 4.4 \cdot 10^{-30}\mathrm{s/Gpc}\), or pulsar timing band \(\tilde{k}/(2\pi) \approx 10^{-8}\mathrm{Hz}\) with \(H_0^3/\tilde{k}^3 \approx 4.4 \cdot 10^{-15}\mathrm{s/Gpc}\)\,.

The magnitude of both effects, the phase shift and the time of arrival difference, depend on a product between two cosmological parameters, the curvature density parameter $\Omega$ and the Hubble parameter $H_0$, and the 1PNGR parameter $\zeta$. Hence, to predict the precise magnitude of gravitational wave birefringence from the theory independent measurements of these parameters are necessary. The other way around, an observation or non-observation of gravitational wave birefringence sets an lower or upper bound on the product of these parameters. To determine the parameters individually, an additional independent measurement, e.g., of the Hubble parameter $H_0$, as well as the source parameters \(\tilde{d}\) and \(\tilde{k}\), is necessary.

\section{Conclusion}
We have calculated the equations of motion and dispersion relation for tensor perturbations, and hence the propagation of gravitational waves, on the most general homogeneous and isotropic background geometry in the 1PNGR class of teleparallel gravity theories. We have found that for one of the two branches of cosmological backgrounds, known as the axial branch, which corresponds to a positive spatial curvature parameter of the FLRW metric, gravitational wave birefringence occurs, i.e., the two polarization states of tensor perturbations propagate at a different speed.

For the other branch, known as the vector branch and representing negative spatial curvature, no gravitational birefringence occurs. Focusing on the axial branch, we have calculated the phase and group velocities for both polarizations, and derived two observable quantities from the difference between the two polarization states: the phase shift and the arrival time difference of a gravitational wave signal. We found that both observables have a distinct dependence on a product between the spatial curvature parameter \(\Omega\), the Hubble parameter $H_0$ and the free parameter \(\zeta\) in the action, and that the time delay has an additional dependence on the wave number $k$. In principle they could be determined by measuring these observables. Further, different theories of gravity can be distinguished by studying the dependence of the birefringence effect on the wave number $k$, which turns out to follow a different power law, e.g., in non-commutative, Chern-Simons or other parity-violating gravity theories.

To estimate the magnitude of the observable effects, we transformed our result to physical units and assumed a single gravitational wave signal emitted from a distant source. We find that the phase shift is independent of the frequency of the wave, and that it is proportional to a factor which is of order unity for sources in the early universe. For the arrival time difference we find a strong suppression by a factor relating the period of the wave to the Hubble time, which is several orders of magnitude beyond any detector sensitivity. Thus we conclude, that the prediction of GW birefringence and dispersion from 1PNGR can in principle be used to find constraints on the parameter $\zeta$ and the spatial curvature parameter \(\Omega\) in the future, however on the basis of the current detector sensitivity no reasonable constraint on these parameters can be placed.

We finally remark that while we restricted the analysis in this letter to 1PNGR, we expect a similar effect to be present also in other teleparallel theories which have a modified coupling of the axial torsion. It is up to further studies to show whether in these theories the effect is stronger and thus more likely to be detectable.

\section*{Acknowledgments}
M.H. acknowledges support by the Estonian Ministry for Education and Science through the Personal Research Funding Grants PRG356, as well as the European Regional Development Fund through the Center of Excellence TK133 ``The Dark Side of the Universe''. C.P. was funded by the Deutsche Forschungsgemeinschaft (DFG, German Research Foundation) - Project Number 420243324 and acknowledges support from the DFG funded cluster of excellence Quantum Frontiers. M.H. and C.P. acknowledge networking support by the COST Action CA18108.

\bibliographystyle{elsarticle-num}
\bibliography{cosmopert}
\end{document}